\documentclass[conference]{IEEEtran}

\usepackage{graphicx}
\graphicspath{{./fig/}}
\DeclareGraphicsExtensions{.pdf .jpg .eps}

\usepackage[english]{babel}
\usepackage{t1enc}
\usepackage[T1]{fontenc}
\usepackage{booktabs}

\usepackage{cite}
\usepackage[hyphens]{url}
\expandafter\def\expandafter\UrlBreaks\expandafter{\UrlBreaks%  save the current one
  \do\a\do\b\do\c\do\d\do\e\do\f\do\g\do\h\do\i\do\j%
  \do\k\do\l\do\m\do\n\do\o\do\p\do\q\do\r\do\s\do\t%
  \do\u\do\v\do\w\do\x\do\y\do\z\do\A\do\B\do\C\do\D%
  \do\E\do\F\do\G\do\H\do\I\do\J\do\K\do\L\do\M\do\N%
  \do\O\do\P\do\Q\do\R\do\S\do\T\do\U\do\V\do\W\do\X%
  \do\Y\do\Z}

\usepackage[multiple]{footmisc}

\usepackage{footnote}
\usepackage{multirow}
\usepackage{amsmath}
\usepackage{listings}

\usepackage{xcolor}

\colorlet{punct}{red!60!black}
\definecolor{background}{HTML}{EEEEEE}
\definecolor{delim}{RGB}{20,105,176}
\colorlet{numb}{magenta!60!black}

\lstdefinelanguage{json}{
    basicstyle=\normalfont\ttfamily,
    numbers=left,
    numberstyle=\scriptsize,
    stepnumber=1,
    numbersep=8pt,
    showstringspaces=false,
    breaklines=true,
    frame=lines,
    literate=
     *{0}{{{\color{numb}0}}}{1}
      {1}{{{\color{numb}1}}}{1}
      {2}{{{\color{numb}2}}}{1}
      {3}{{{\color{numb}3}}}{1}
      {4}{{{\color{numb}4}}}{1}
      {5}{{{\color{numb}5}}}{1}
      {6}{{{\color{numb}6}}}{1}
      {7}{{{\color{numb}7}}}{1}
      {8}{{{\color{numb}8}}}{1}
      {9}{{{\color{numb}9}}}{1}
      {:}{{{\color{punct}{:}}}}{1}
      {,}{{{\color{punct}{,}}}}{1}
      {\{}{{{\color{delim}{\{}}}}{1}
      {\}}{{{\color{delim}{\}}}}}{1}
      {[}{{{\color{delim}{[}}}}{1}
      {]}{{{\color{delim}{]}}}}{1},
}

\usepackage{url}
\usepackage[backgroundcolor=blue!10,bordercolor=gray]{todonotes}

\begin{document}

\title{Static JavaScript Call Graphs: a Comparative Study}

\author{\IEEEauthorblockN{G\'abor Antal\IEEEauthorrefmark{1},
P\'{e}ter Heged\H{u}s\IEEEauthorrefmark{2},
Zolt\'an T\'oth\IEEEauthorrefmark{1},
Rudolf Ferenc\IEEEauthorrefmark{1}, and
Tibor Gyim\'othy\IEEEauthorrefmark{1}\IEEEauthorrefmark{2}}
\IEEEauthorblockA{\IEEEauthorrefmark{1}Department of Software Engineering, University of Szeged, Hungary\\
E-mail:\{antal~|~zizo~|~ferenc\}@inf.u-szeged.hu}
\IEEEauthorblockA{\IEEEauthorrefmark{2}MTA-SZTE Research Group on Artificial Intelligence, Szeged, Hungary\\
E-mail: \{hpeter~|~gyimothy\}@inf.u-szeged.hu}
}

\maketitle

\begin{abstract}
The popularity and wide adoption of JavaScript both at the client and server side makes its code analysis more important than ever before.
Most of the algorithms for vulnerability analysis, coding issue detection, or type inference rely on the call graph representation of the underlying program.
Despite some obvious advantages of dynamic analysis, static algorithms should also be considered for call graph construction as they do not require extensive test beds for programs and their costly execution and tracing.

In this paper, we systematically compare five widely adopted static algorithms -- implemented by the npm call graph, IBM WALA, Google Closure Compiler, Approximate Call Graph, and Type Analyzer for JavaScript tools -- for building JavaScript call graphs on 26 WebKit SunSpider benchmark programs and 6 real-world Node.js modules.
We provide a performance analysis as well as a quantitative and qualitative evaluation of the results.

We found that there was a relatively large intersection of the found call edges among the algorithms, which proved to be 100\% precise.
However, most of the tools found edges that were missed by all others.
ACG had the highest precision followed immediately by TAJS, but ACG found significantly more call edges.
As for the combination of tools, ACG and TAJS together covered 99\% of the found true edges by all algorithms, while maintaining a precision as high as 98\%.
Only two of the tools were able to analyze up-to-date multi-file Node.js modules due to incomplete language features support.
They agreed on almost 60\% of the call edges, but each of them found valid edges that the other missed.

%az alábbi két mondatot semmi sem támasztja alá, mivel egyrészt nincs recall infónk, másrészt nincs még dinamikus gráfunk. De szerintem enélkül kerek az abstract.
%We propose that a combination of fast static algorithms used together could be a viable alternative for building a decent base call graph.
%However, we anticipate that best results could be achieved by combining static and dynamic approaches for JavaScript call graph construction.
\end{abstract}


\vspace{-5pt}
\section{Introduction}\label{sec:introduction}

According to GitHub statistics~\cite{GitHub_stat2} JavaScript is one of the most rising languages in years, and it seems that it will continue to dominate in 2018.
It had the most pull requests in 2017 and 2016 (in GitHub projects).
Each year, the TIOBE Index selects the fastest growing programming language and distinguishes it with the ``Language of the Year'' award.
In 2014, JavaScript was the winner of this award.
%Moreover, JavaScript takes a prominent place in the list of open software developer jobs~\cite{Dev_jobs2}.

%JavaScript is also one of the most popular programming languages in startup companies~\cite{Startup1}.
%JavaScript might be a reasonable choice because it can be used both on the server and client side, has a really large library support, and there are lots of people who can help if someone faces a problem.
%ez nem logikus, kiveszem: There are also many developers available, which further increases the popularity of the JavaScript language.

Due to its increasing popularity, lots of projects use JavaScript as their core programming language for both server and client side modules.
Therefore, static code analysis of JavaScript programs became a very important topic as well.
Many of the code analysis tools rely on the call graph representation of the program.
A call graph contains nodes that represent functions of the program and edges between nodes if there exists at least one function call between the corresponding functions.
With the help of this program representation various quality and security issues can be detected in JavaScript programs, for example, it can be used to detect functions that are never called or as a visual representation which makes understanding the code easier.
We can use call graphs to examine whether the correct number of arguments is passed to function calls or as a basis for further analysis, for example, a full interprocedural control flow graph (ICFG) can be built upon the call graph.
With the help of the control flow graphs, various type analysis algorithms can be performed~\cite{jensen2009type, feldthaus_acg, madsen2013practical, pradel2015typedevil}.
What is more, this program representation is useful in other areas of research as well, for example, in mutation testing~\cite{mirshokraie2013efficient}, automated refactoring~\cite{feldthaus2011tool}, or defect prediction~\cite{bhattacharya2012graph}.

Being such a fundamental data structure, the precision of call graphs determines the precision of the code analysis algorithms that rely on them.
Creating precise call graphs for JavaScript that is an inherently dynamic, type-free and asynchronous language is quite a big challenge.
Static approaches have the obvious disadvantage of missing dynamic call edges coming from the non-trivial usages of $eval()$, $bind()$, or $apply()$ (i.e. reflection).
Moreover, they might be too conservative, meaning that they can recognize statically valid edges, which are never realized for any inputs in practice.
However, they are fast and efficient compared to dynamic analysis techniques and do not require any testbed for the program under analysis.

Therefore, the state-of-the-art static call graph construction algorithms for JavaScript should not be neglected and we need deeper understanding about their performance, capabilities, and limitations.
%We believe that the optimal code analysis strategy for JavaScript programs would be hybrid, which combines the power of static analysis and dynamic runtime execution traces.
%As a vital step to reach this goal, 
In this paper we present and compare some well-known and widely used static analysis based call graph building approaches.
We compare five different tools -- npm call graph, IBM WALA, Google Closure Compiler, ACG (Approximate Call Graph), and TAJS (Type Analyzer for JavaScript) -- quantitatively, to find out how many different calls are detected by the individual tools.
We also compare the results qualitatively, meaning that we match and validate the found call edges and analyze the differences.
Lastly, we report runtime and memory usage data to be able to assess the usability of the tools on real-world programs.

%We found that there is a high correlation between the found node and edge counts among the tools, however, they are far from being identical.
We found that there are variances in the number, precision and type of call edges that individual tools report.
However, there were considerably large intersections of the reported edges.
% among the tools (about one third in case of five tools on SunSpider benchmark and two thirds for two tools on the 6 Node.js modules).
Based on a manual evaluation of 348 call edges, we found that ACG had the highest precision, above 99\% of the found edges were true calls.
At the same time, ACG had the highest recall on the union of all true edges found by the five tools, it found more than 90\% of the edges.
Nonetheless, three other tools (WALA, Closure, npm call graph) found true positive edges that were missed by all the other tools.
TAJS did not find any unique edges, however it achieved a precision of 98\% (i.e. comparable to ACG).
%For example, Closure was the only tool recognizing recursive calls, WALA tracking the calls realized by calling a function passed as a parameter, and ACG to follow complex control paths to extract possible function calls.
We also examined the tools in combination and saw that ACG, Closure, and TAJS together found all the true edges, but they also introduced a lot of false ones, their combined precision was only slightly above 83\%.

In terms of running time performances, results heavily depend on the size and complexity of the inputs, but Closure and TAJS excel in this respect.
%produced the best one noticeable thing is that Closure runs much faster than the others when the input file is large but simple (there are no callbacks or callback calls).
%But when the input file is complex (there are lots of callbacks and callback calls), TAJS was the fastest.
From the perspective of memory consumption, for realistic input sizes ACG and Closure overtopped all the other tools.
%in the case of simple, small inputs, ACG was the winner.
%But when there is a complex input, both ACG and Closure Compiler are good choices.
For very large inputs (i.e. in the range of a million lines of code), only Closure Compiler, TAJS and ACG were able to perform practically efficient code analysis.

To summarize, the main contributions of our paper are:
\begin{itemize}
	\item The evaluation of capabilities and performances of five widely adopted static JavaScript call graph extraction tools.
	\item The quantitative and qualitative comparison of the tool results on 26 benchmark programs and 6 real-world Node.js modules.
	\item A manually validated dataset of call edges found by these tools on the 26 benchmark programs.
\end{itemize}

The rest of the paper is organized as follows.
In Section~\ref{sec:related}, we list the related literature and compare our work to them.
Section~\ref{sec:methodology} describes the tool selection and comparison methodology we applied.
In Section~\ref{sec:results}, we present the results of our quantitative, qualitative, and performance analysis of the tools.
We list the possible threats of the analysis in Section~\ref{sec:threats} and conclude the paper in Section~\ref{sec:conclusions}.
\section{Related Work}\label{sec:related}

%A bevezető blokkból ki lehet szedni néhány alkalmazást ha kell
Using call graphs for program analysis is a well-established and mature technique.
The first papers dealing with call graphs date back to the 1970's~\cite{allen1974interprocedural, Graham:1982:GCG:872726.806987}.
The literature is full with different studies built upon the use of call graphs.
Clustering call graphs can have advantages in malware classification~\cite{kinable2011malware}, they can help localizing software faults~\cite{eichinger2008mining}, not to mention the usefulness of call graphs in debugging~\cite{rao2013debugging}.

Call graphs can be divided into two subgroups based on the used method to construct them.
These two groups are dynamic~\cite{xie2002empirical} and static call graphs~\cite{murphy1998empirical}.
%innen is kiszedhet az egyik hivatkozás
Dynamic call graphs are obtained by running the program and collecting runtime information about the interprocedural flow~\cite{eichinger2010localizing}.
Techniques such as instrumenting the source code can be used for dynamic call graph creation~\cite{dmitriev2004profiling}.
In contrast, in case of static call graphs there is no need to run the program, it is produced as a result of static analysis of a program.
%Static call graphs often include extra edges since in these cases it looks possible to call a given function/method at a point, however it can happen that an actual call never executes due to a condition.
%Static call graphs can be constructed in any case from the source code even if it could not be run.
Different analysis techniques are often combined to obtain a hybrid solution which guarantees a more precise call graph, thus a more precise analysis~\cite{eisenbarth2001aiding}.

With the spread of scripting languages such as Python and JavaScript the need for analyzing programs written in these languages also increased~\cite{feldthaus_acg}.
However, constructing precise static call graphs for dynamic scripting languages is a very hard task which is not fully solved yet.
The \emph{eval()}, \emph{apply()} and \emph{bind()} constructions of the language make it especially hard to analyze the code statically.
However, there are several approaches to construct such static call graphs for JavaScript with varying success~\cite{feldthaus_acg, bolin2010closure, fink2012wala}.
Constructed call graphs are often limited, and none of the studies deal fully with EcmaScript 6 since the standard was released in 2015.

%Wei and Ryder presented blended taint analysis for JavaScript which uses a combined static-dynamic analysis~\cite{wei2013practical}.
%By applying dynamic analysis they could collect information for even the constructions which are hard to analyze statically.
%Dynamic results (execution traces) are propagated to a static infrastructure which embeds a call graph builder as well.
%This call graph builder module makes use of the dynamically identified calls.
%However, in case of pure static analysis they wrapped the WALA tool to construct a static call graph.
%We also included WALA in our comparison study.

Feldthaus et al. presented an approximation method to construct a call graph~\cite{feldthaus_acg} by which a scalable JavaScript IDE support could be guaranteed.
Madsen et al. focused on the problems induced by libraries used in the project~\cite{madsen2013practical}.
They used pointer analysis and a novel use analysis to enhance scalability and precision.
In our study, we only deal with static call graphs for JavaScript and do not propose a new algorithm, but rather evaluate and compare existing approaches.

In his thesis~\cite{dijkstra2014evaluation}, Dijkstra evaluates various static JavaScript call graph building algorithms.
This work is very similar to our comparative study, however, it was published in 2014 and a lot has happened since then in this research area.
Moreover, while Dijkstra focused on the evaluation of the various conceptual algorithms implemented by himself in Rascal, our focus is on comparing mature and state-of-the-art tool implementations on these algorithms ready to be applied in practice.

There are also works with a goal to create a framework for comparing call graph construction algorithms~\cite{lhotak2007comparing, ali2012application}.
However, these are done for algorithms written in Java and~C.
Call graphs are often used for preliminary analysis to determine whether an optimization can be done on the code or not.
Unfortunately, as they are specific to Java and C languages, we could not use these frameworks as is for comparing JavaScript call graphs.

%nem igazan ide valo... Our long term plan is to design a precise, more flexible, hybrid state-of-the-art call graph construction algorithm for JavaScript with support for ES 6 and later standards.
%As a first step, in this comparative study we quantitatively and qualitatively evaluate existing static call graph construction algorithms for JavaScript.

%Comparing call graphs~\cite{lhotak2007comparing}\todo{ha kell valahova}
\vspace{-10pt}
\section{Methodology}
\label{sec:methodology}

\subsection{Overview of the study process}
\label{sec:methodology-overview}

\begin{figure}
\includegraphics[width=\columnwidth]{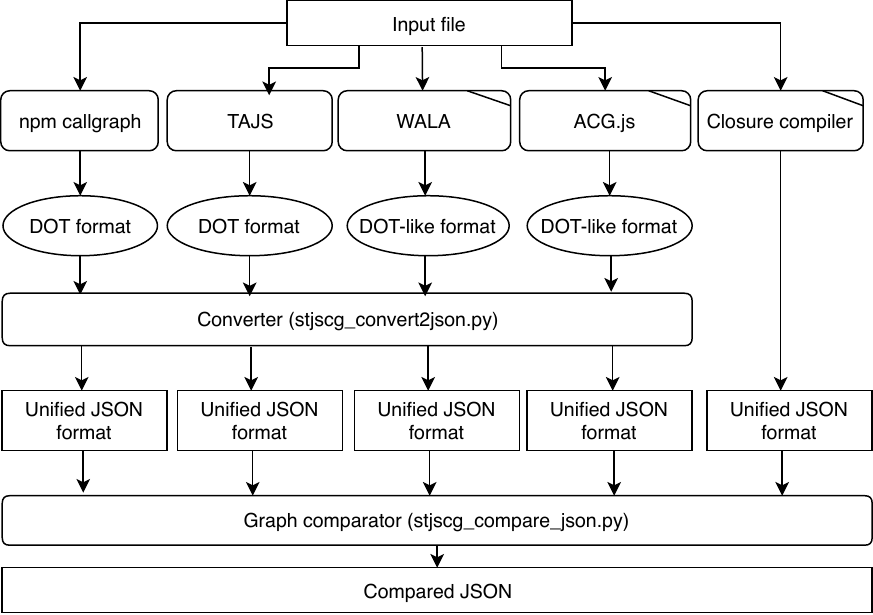}
\vspace{-17pt}
\caption{Methodology overview}
\label{fig:methodology}
\vspace{-16pt}
\end{figure}

Figure~\ref{fig:methodology} displays the high-level overview of the applied external and self-developed software components we used in our comparative study.
We run each of the selected tools (Section~\ref{sec:tools}) on the test input files (Section~\ref{sec:subjects}).
As can be seen, we needed to patch some of the tools (marked with \textbackslash) for various reasons (see Section~\ref{sec:tools}), but mainly to extract and dump the call graphs built in the memory of the programs (all the modification patches are available in the online appendix package\footnote{\vspace{-5pt}\url{http://www.inf.u-szeged.hu/~ferenc/papers/StaticJSCallGraphs/}}).
Next, we collected the produced outputs of the tools and ran our data conversion scripts to transform each call graph to a unified, JSON based format we defined (Section~\ref{sec:output}).
The only exception was Closure, where we implemented the call graph extraction to the JSON format right into the patch extracting the inner-built call graph, because there was no public option for outputting it, thus it was easier to dump the data right into the unified JSON format.
In all other cases we built a custom data parser script that was able to read the output of the tools and produce an equivalent of it in our JSON format.
From the individual JSON outputs of the tool results, we created a merged JSON with the same structure using our graph comparison tool (Section~\ref{sec:graph-compare}).
This merged JSON contains all the nodes and edges found by either of the tools, with an added attribute listing all the tool ids that found that particular node or edge.
We ran our analysis and calculated all the statistics on these individual and merged JSON files (all the produced JSON outputs are part of the online appendix package).

\vspace{-3pt}
\subsection{Call graph extraction tools}
\label{sec:tools}
In this section, we present the tools we took into account in our comparative study.
We examined tools that: i) are able to create a function call graph from a JavaScript program, ii) are free and open-source, and iii) are adopted in practice.

It is important to note that in this study we work with call graphs, where:
\begin{itemize}
	\item Each node represents a function in the program (identified by the file name, line and column number of the function declaration),
	\item An edge between two nodes is directed and represents a statically possible call from one function to another (i.e. function \emph{f()} may call function \emph{g()}),
	\item There might be only zero or one edge between two nodes, so we track only if a call is possible from one function to another, but omit its multiplicity (i.e. we do not count at how many call sites calls may happen). This is because not all of the tools are able to find multiple calls and we wanted to stick to the most basic definition of the static call graph anyway.
\end{itemize}

Based on these criteria, we selected the following five tools for our comparative study (see Table~\ref{table:tool_comparison} for an overview).

\begin{table*}[ht!]
    \centering
		\caption{Comparison of the used tools (as of 16th July, 2018)}
		\scriptsize
    \begin{tabular}{|l|l|r|r|r|r|r|c|l|}
        \hline
        Tool name        & Repository & Lang. & Size   & Commits & Last commit & Contri- & Issues        & ECMAScript \\
				                 &            &          & (SLOC) &         &             & butors    & (open/closed) & compatibility \\
				\hline
        WALA             & \url{https://github.com/wala/WALA} & Java & 232,594      & 5,845    & 06/11/2018 & 25 & 151 (74/77)          & ES5             \\ \hline
        Closure compiler & \url{https://github.com/google/closure-compiler} & Java & 398,959      & 12,525   & 06/16/2018 & 373 & 2163 (796/1367)      & ES6 (partial)  \\ \hline
        ACG              & \url{https://github.com/cwi-swat/javascript-call-graph} & JS & 120,531      & 193     & 10/28/2014 & 3 & 7 (1/6)              & ES5             \\ \hline
        npm callgraph    & \url{https://github.com/gunar/callgraph} & JS & 207         & 30      & 03/14/2017 & 2 & 16 (6/10)            & ES6 (partial)  \\ \hline
				TAJS             & \url{https://github.com/cs-au-dk/TAJS} & Java&   53,228       & 16      & 01/04/2018 & 1 & 10 (6/4)            & ES5 (partial)   \\ \hline
    \end{tabular}		
    \label{table:tool_comparison}
		\vspace{-15pt}
\end{table*}

\subsubsection{WALA}
WALA~\cite{fink2012wala} is a complete framework for both static and dynamic program analysis for Java.
It also has a JavaScript front-end, which is built on Mozilla's Rhino~\cite{rhinojavascript} parser.
In this study, we used only one of its main features, which is static analysis, call graph construction in particular.

In order to have the output that suits our needs, we had to create a driver which serializes the built call graph.
For this, we used an already existing version of the call graph serializer found in the official WALA repository (\emph{CallGraph2JSON.java}),
As a first step, we converted the actual call graph to a simple DOT format then we used our converter script to transform this into the final JSON file.
WALA produced multiple edges between two functions if there were multiple call sites within the caller function.
Since our definition of call graph allows at most one edge between two functions in one direction, we modified the serializer to filter the edges and merge them if necessary.
We had to handle the special case when the call site was in the global scope as in this case there was no explicit caller method.
In such cases we applied the common practice followed by other tools as well and introduced an artificial ``toplevel'' node as the source of these edges.

WALA itself is written entirely in Java, its
%, but there are also complementary tools implemented in JavaScript~\cite{JS_WALA}.
%However, that project is barely used and even barely developed, thus we chose the main Java implementation.
main repository is under active development, mostly by the IBM T.J. Watson Research Center.
WALA was used in over 60 publications~\cite{walapublications} since 2003.
%The repository has more than 5,400 commits from~25 contributors and it has 74 opened and~77 closed issues.

\subsubsection{Closure Compiler}

The Closure Compiler~\cite{bolin2010closure} is a real JavaScript compiler, which works as other compilers.
But in the case of Closure Compiler it compiles JavaScript to a better JavaScript: it parses and analyzes JavaScript programs, removes dead code, rewrites and compresses the code.
It also checks common JavaScript mistakes.

It builds a call graph data structure for internal use only by other algorithms.
Therefore, we had to modify the available source code and provide a call graph JSON dump function.
Closure Compiler contains the inclusion of the artificial root node by default to represent calls realized from the global scope.
The JSON writer filters any duplicate edges (Closure keeps track of various call sites) to provide an appropriate JSON output used for comparison (see Section~\ref{sec:output}).

The Closure Compiler itself is written entirely in Java and is actively developed by Google.
%The repository has more than~12,000 commits from 373 contributors and it also has 797 open and 1365 closed issues.

\subsubsection{ACG}

ACG (Approximate Call Graph) implements a field-based call graph construction algorithm~\cite{feldthaus_acg} for JavaScript.
The call graph constructor can be run in two basic modes, pessimistic and optimistic, which differ in how interprocedural flows are handled.
%There are several strategies that use the same intraprocedural flow graph, in which properties are only identified by name; thus, like-named properties of different objects are conflated; this can lead to imprecise call graphs.
%Dynamic property reads and writes are ignored, as are reflective calls; thus, the call graphs are intrinsically incomplete.
In our study, we used the default \texttt{ONESHOT} pessimistic strategy for call graph construction.

For ACG, we had to implement the introduction of an artificial root edge (i.e. ``toplevel'') and filtering of multiple edges as ACG also tracks and reports edges connected to individual call sites.
Moreover, ACG reported only the line numbers of functions in its output, which we had to extend with the column information.
All these modifications are available in one single patch.

%Since the original repository\footnote{\vspace{-20pt}\url{https://github.com/xiemaisi/acg.js}} is abandoned, we selected a fork from this repository.
As there are several forks of the original repository available currently, we had to check all of them and select the one which is the most mature among these forks.
The selected one was created by the CWI group from Amsterdam.
%This repository has 193 commits from~3 authors in the release branch.

\subsubsection{The npm callgraph module}

Npm callgraph is a small npm module to create call graphs from JavaScript code developed by Gunar C. Gessner.
It uses UglifyJS2~\cite{bazon2016uglifyjs} to parse JavaScript code.
Despite its small size and few commits, quite a lot of people use it, it has more than 1300 downloads.
%133 people installed the tool just in February 2018.

%Even though the author does not update the code frequently, he quickly reacts to newly opened issues.
%The original repository has 30 commits from 2 authors and it has 6 opened and 10 closed issues.
%During the analysis, we ran into \texttt{TypeError} issues several times.
%A simple null check solved the problem and we never encountered this kind of error again.
%We created a fix for this issue and proposed a pull request to the repository, which was already approved and merged to the master branch.\footnote{\url{https://github.com/gunar/callgraph/commit/36bab6a0a437c04c2518ae5c4b108791c706eb07}}
%We are preparing several patches (from those modifications which were necessary for us, as it can be useful for others too) to create pull requests for the original tool.

\subsubsection{TAJS}

Type Analyzer for JavaScript~\cite{tajs2009} is a dataflow analysis tool for JavaScript that infers type information and call graphs.
It is copyrighted to Aarhus University.

The proposed algorithm is implemented as a Java system that is actively maintained since the publication of the original concept.
%Nonetheless, the code repository has only 16 commits from 1 contributor, 6 open and 4 closed issues.
However, we suspect that this is only an external mirror of an internal repository that is synchronized periodically.
%As we browsed its source code, we found comments that refers to non-existent issues (as on GitHub, there are only 10 issues, while we saw issue ticket \#332 in the source code).
%We assume that the GitHub repository is only a mirror, while the original repository is private.
It was not necessary to modify the source code of TAJS as it provides a command line option for dumping call graphs into a DOT format that we were able to parse and convert into our unified JSON format.

\textbf{Other considered tools.}
% The selection criteria in §I-B for example do not mention anything about handling multiple files or JavaScript versions.
%
There are of course other candidate tools which could have been involved in this study.
We found lots of commercial and/or closed-source programs, like SAP HANA.
However, we focused on open-source programs, which are easy to access and even modify or customize to fit our needs.
They are also widely adopted by research and industry.

In our evaluation study we only dealt with tools directly supporting call graph building either internally or as a public feature.
Thus, we were forced to left out some great JavaScript static analysis tools that do not support call graph extraction directly. 
One such tool was the open-source Flow~\cite{Flow} developed by Facebook, a very popular static code analysis tool for type checking JavaScript.
%Flow performs both data flow and control flow analysis to infer types.
%It is modular and fast, suitable to analyze even millions of lines of code.
%Flow was written in OCaml (Objective Caml) which is a multi-paradigm, general-purpose programming language.
Unfortunately, Flow does not provide a public API for obtaining the built call graph or a control flow graph.
As such, we would have been required to implement our own algorithms above the internal control flow data structure, which would introduce a threat to the validity of this study.
Our primary goal in this work was to empirically compare existing call graph extraction algorithms, not to upgrade all tools to achieve call graph extraction.
%It is a challanging task to understand this huge codebase (\textasciitilde240k lines of code) and implement our desired call graph extractor.

%Being an exotic language in which we lacked appropriate expertise, we could not manage to extract the in-built call graph representation from the tool.
%Moreover, lots of modifications would have been needed to obtain a complete, static call graph.
%We even tried to contact the developers of Flow on the official IRC channel multiple times, but so far we have not got any response.

%Type Analyzer for JavaScript (TAJS)~\cite{tajs2009} by Aarhus University also looked promising.
%The main disadvantage with TAJS was that it supports only EcmaScript 3 fully and EcmaScript 5 only partially.
%Also, it targets JavaScript running in a browser, which was not our primary focus.
%Moreover, it depends on Closure Compiler by Google, which we included in our study.
%Taking everything into consideration, we decided to drop it from the list. 

Other relevant tools we examined were JSAI (JavaScript Abstract Interpreter)~\cite{Kashyap:2014:JSA:2635868.2635904} and SAFE (Scalable Analysis Framework for EcmaScript)~\cite{lee2012safe}, both build an intermediate abstract representation from JavaScript to further perform an analysis on.
It is true that they calculate control and data flow structures, but they specifically utilize them for type inference.
None of them support the extraction/export of call graphs, hence we were unable to include them in our evaluation study.

%Flow, JSAI, and SAFE are all great tools but the inner representations are difficult to modify in order to export the call graphs.
%In case of Closure Compiler we could make a small patch to fulfill our needs.
%In case of Flow, JSAI, and SAFE it demands a lot more work to obtain the same.

The tool \texttt{code2flow}~\cite{code2flow} also looked like a great choice, but since it is abandoned officially without any follow-up forks, we excluded it from our list.
We note that the original repository of ACG was also abandoned, however it has several active forks on GitHub.

Another reason we dropped possible tools from comparison was immaturity, which means that the given project had one contributor and there was only a very short development period before the project was left abandoned.
These tools also lacked documentation, thus their usability was poor.
We did not take into account \texttt{JavaScript Explorer Callgraph}~\cite{JEC} due to this reason.
Furthermore, we also left out \texttt{callgraphjs}~\cite{cgjs} since this project contains only supporting material for ACG.

%%%%%%%%%%%%%%%%%%%%%%%%%%%%%%%%%%%%%%%%%%%%%%%%

\vspace{-4pt}
\subsection{Comparison subjects}
\label{sec:subjects}
\vspace{-1pt}
To perform a deep comparison of the tools, we identified three test input groups.
%The first group consists of real-life, single file, ``bare'' JavaScript examples.
%These kind of programs are quite small but they test both the used parser and the call graph building algorithm.
%And of course, they are relatively small files that can be manually checked and analyzed.

%The second group of test inputs are those real-life programs that contain many separate JavaScript source files and use advanced EcmaScript 6 features, like module exports or external dependencies (i.e. use the \emph{require} keyword).
%These kind of programs typically target the Node.js platform.

%The last group of test subjects is the generated, large inputs.
%This group contains generated programs with various code sizes, from few hundreds to a million lines of code.
%These kinds of programs are useful for stress testing the tools and the underlying algorithms and to measure performance.

\subsubsection{Real-world, single file examples}
The first group consists of real-world, single file, ``bare'' JavaScript examples.
For this, we chose the SunSpider benchmark of the WebKit browser engine~\cite{SunSpider}, which is extensively used in other works as well.
The benchmark programs are created to test the JavaScript engine built into WebKit.
Therefore, these programs contain varying complexity code with many different types of functions and calls, but all in one single JavaScript source file.
These properties make them excellent choice for our real-world, single-file test subjects.
Moreover, all the programs are of manageable size, thus we could easily check and analyze the calls manually.

\subsubsection{Real-world, multi-file Node.js examples}
\label{sec:reallife_node}
To test the handling of modern, EcmaScript 6 and Node.js features (like module exports or external dependencies, i.e. the \emph{require} keyword) and inter-file dependencies, we collected six popular Node.js modules from GitHub.
Our selection criteria included the following: the module should contain multiple js source files, it should have an extensive test suite with at least 75\% code coverage and be used by at least 100 other Node.js modules.
The requirements for test coverage comes from our mid-term research goal.
We would like to repeat the presented comparative study extended with dynamic call graph extraction algorithms that typically require an existing test suite for programs under analysis.
The details of the chosen Node.js modules are summarized in Table~\ref{tab:node-examples}.

% Table generated by Excel2LaTeX from sheet 'Munka2'
\begin{table}[htbp]
  \centering
	\vspace{-8pt}
  \caption{Selected Node.js modules for test}
	\scriptsize
	\vspace{-7pt}
    \begin{tabular}{|l|l|r|}
    \hline
    Program & Repository & \multicolumn{1}{l|}{Size (SLOC)} \\
    \hline
    debug & \url{https://github.com/visionmedia/debug} & 442 \\
    \hline
    doctrine & \url{https://github.com/eslint/doctrine} & 5,109 \\
    \hline
    express & \url{https://github.com/expressjs/express} & 11,673 \\
    \hline
    jshint & \url{https://github.com/jshint/jshint} & 68,411 \\
    \hline
    passport & \url{https://github.com/jaredhanson/passport} & 6,173 \\
    \hline
    request & \url{https://github.com/request/request} & 9,469 \\
    \hline
    \end{tabular}%
  \label{tab:node-examples}%
	\vspace{-8pt}
\end{table}%

\subsubsection{Generated large examples}
In order to stress test the selected tools and measure their performances, we needed some really large programs.
However, we were unable to find large enough open-source programs, which would use only those language features that all of the tools recognize.
%The first step was to search open source projects we can feed into the tools we selected.
%However, we failed to find open source JavaScript programs with source code size bigger than 1 million lines of code.
%Furthermore, these open source programs often use various features from both ECMAScript 5 and 6. But unfortunately not all of the tools were capable to parse ES6.
Therefore, we decided to generate JavaScript programs that conform to the ECMAScript 5 standard as it is the highest standard all the selected tools support.
%The generated program have various sizes and have lots of calls in it.

We defined two categories of such generated inputs.
The programs in the \textit{simple} category contain simple function calls with some random statements (variable declarations, object creation and object property access, for loops, while loops or return statements).
They are pretty straightforward without any complex control flows, but their sizes range from moderate to very large.
The programs in the \textit{complex} category also contain numerous function calls but extended with functions with parameters, callback functions, function expressions, loops, and simple logging statements.
These programs meant to test the performances of tools when parsing complex control flows.
The code generation was performed automatically, with custom made Python scripts.

%, we created two simple tools: the first one is able to generate a given number of functions.
%The function blocks may contain any of the following items randomly: variable declarations, object creation and object property access, for loops (for sorting an array), while loop (creating a %single string from array items) or return statement.
%The script automatically creates functions with easily identifiable names (e.g. function48), and the calls are randomly placed inside the functions (where only the already defined functions can %be called).
%Its output is straightforward and it can be easily analyzed even with a simple textual analysis.
%So we decided to create a tool which can generate JavaScript more similar to the real world. Our second tool can generate: functions with parameters; callback functions; function expressions; %loops; simple logging statements, and calls between the functions.
%Calls are made even from loops, and callbacks.
%(the scripts are available in our online appendix package).

We generated three files in the simple and two in the complex category (the exact properties of these programs are shown in Table~\ref{table:stats_randomjs}).
After the generation, we used the Esprima Syntax Validator\footnote{\vspace{-20pt}\url{http://esprima.org/demo/validate.html}} to validate our files thus they are valid JavaScript programs and can be parsed with any ECMAScript 5 compatible JavaScript front-end.

\begin{table}[htb]
\vspace{-8pt}
\centering
\caption{Properties of the random generated JavaScript files}
\vspace{-4pt}
\scriptsize
\begin{tabular}{|l|l|r|r|r|}
\hline
Type                     & File       & Code lines & Nodes   & Edges       \\ \hline
\multirow{3}{*}{Simple}  & s\_small.js   & 68,741     & 1000    & 49,286      \\ \cline{2-5}
                         & s\_medium.js  & 382,536    & 2,600   & 331,267     \\ \cline{2-5}
                         & s\_large.js   & 1,321,088  & 5,000   & 1,224,251   \\ \hline
\multirow{2}{*}{Complex} & c\_medium.js  & 28,544     & 400     & 3,000       \\ \cline{2-5}
                         & c\_large.js   & 413,099    & 1,000   & 50,000      \\ \hline
\end{tabular}
\label{table:stats_randomjs}
\vspace{-12pt}
\end{table}

%We created two categories to test the performance of each tool.
%The \textit{simple} category contains simple function calls with some random statements.
%The \textit{complex} category also contains numerous function calls but extended with random callback functions, calls with parameters, and variable declarations (function expressions are %generated as well).
%The \textit{simple} category holds 3 generated source files varying in size small (\textasciitilde 70,000 lines of code), medium (\textasciitilde 380,000 lines of code) and large (\textasciitilde %1,300,000 lines of code).
%The \textit{complex} category includes 2 source files such that medium has \textasciitilde 28,000 lines of code and large has \textasciitilde 413,000 lines of code.
%The number of nodes and edges are also shown beside the lines of code in Table~\ref{table:stats_randomjs}.

%%%%%%%%%%%%%%%%%%%%%%%%%%%%%%%%%%%%%%%%%%%%%%%%
\subsection{Output format}\label{sec:output}
The different selected tools produce their outputs in different formats by default. %, which would make comparing them a very difficult task.
To solve this problem, we had to process their outputs and convert them into a unified format that can be used for further analysis.
We chose a simple JSON format that contains the list of nodes and edges of a call graph.%demonstrated in Listing~\ref{lst:json}.

%\begin{lstlisting}[caption={JSON output example},language=json,firstnumber=1,label=lst:json,captionpos=b,basicstyle=\scriptsize ,xleftmargin=3.5ex]
%{
%  "nodes": [
%    {
%      "id": 1,
%      "label": "[simple.js]function1()",
%      "pos": "simple.js:10:1"
%    },
%    {
%      "id": 2,
%      "label": "[simple.js]function2()",
%      "pos": "simple.js:18:1"
%    }
%  ],
%  "links": [
%    {
%      "target": 2,
%      "source": 1,
%      "label": "[simple.js]function1() -> [simple.js]function2()"
%    }
%  ]
%}
%\end{lstlisting}
%\vspace{-5pt}

Each node has a unique identifier (id), a label, and source code position information.
The position information clearly identifies a function (i.e. node). %the node, since the identifier can be different for different tools, which would make the comparison impossible.
Each edge connects exactly two of the nodes by their unique ids.

%%%%%%%%%%%%%%%%%%%%%%%%%%%%%%%%%%%%%%%%%%%%%%%%
\vspace{-4pt}
\subsection{Graph comparison}
\label{sec:graph-compare}

The quantitative analysis of the call graphs focuses on the comparison of the number of nodes and edges.
For the qualitative analysis -- inspired by the work of Lhot{\'a}k et. al~\cite{lhotak2007comparing} --, we created a call graph comparison script written in Python.
The script is available in our online appendix package.
The aim of the script is to detect the common edges found by different tools.
%The inputs of the script are the converted unified format JSON outputs (see Figure~\ref{fig:methodology} and Listing~\ref{lst:json}) of the tools and its output is a similar JSON file that contains the union of the found nodes and edges of all tools for each program.
The script decorates each node and edge JSON entry with a list of tool identifiers that found the particular node or edge.
The identification of nodes and edges are done by precise path, line, and column information as JavaScript functions have no names and it would be cumbersome to rely on a unified unique naming scheme anyway.
%, thus our comparison heavily relies on the precision of the line information produced by the tools.
%We chose this strategy as in JavaScript functions have no names and it would be cumbersome to rely on a unified unique naming scheme anyway.
%Moreover, the Chrome V8~\cite{v8} JavaScript engine follows the same strategy.

To ensure the proper comparison, we manually checked the produced path and line information of the evaluated tools.
TAJS reported precise line and column information in its standard DOT output.
In cases of Closure Compiler, WALA, and ACG, we implemented or modified the line information extraction.
% (all the modifications to the tools are available as patches in our online appendix package).
Unfortunately, WALA was able to report only line numbers, but no column information, thus we manually refined the JSON outputs it produced.
In case of npm callgraph, the reported line information was not precise (neither line, nor column information), thus we went through all the cases manually and added them to the produced JSON files.
%The result of the graph comparison on the 26 SunSpider benchmark programs and the~6 Node.js modules are available in the online appendix.

\begin{table*}[ht!]
    \centering
		\caption{SunSpider results}
		\vspace{-5pt}
		\scriptsize
    \begin{tabular}{|l|l|l|l|l|l|l|l|l|l|l|}
        \hline
        & \multicolumn{2}{c|}{npm callgraph} & \multicolumn{2}{c|}{ACG} & \multicolumn{2}{c|}{WALA} & \multicolumn{2}{c|}{Closure Compiler} & \multicolumn{2}{c|}{TAJS} \\ \hline
        file                      & nodes & edges & nodes & edges & nodes & edges & nodes & edges & nodes & edges \\ \hline
        3d-cube										& 15    & 23    & 15    & 22    & 17    & 24    & 15    & 23    & 15    & 23 \\ \hline
				3d-morph 									& 2     & 1     & 2     & 1     & 0     & 0     & 2     & 1     & 2     & 1  \\ \hline
				3d-raytrace 							& 22    & 29    & 28    & 40    & 21    & 22    & 27    & 40    & 28    & 39 \\ \hline
				access-binary-trees 			& 3     & 3     & 4     & 3     & 4     & 5     & 4     & 5     & 4     & 5  \\ \hline
				access-fannkuch 					& 2     & 1     & 2     & 1     & 3     & 2     & 2     & 1     & 2     & 1  \\ \hline
				access-nbody 							& 8     & 11    & 12    & 15    & 8     & 11    & 11    & 14    & 12    & 15 \\ \hline
				access-nsieve 						& 3     & 2     & 3     & 2     & 2     & 1     & 3     & 2     & 3     & 2  \\ \hline
				bitops-3bit-bits-in-byte 	& 2     & 1     & 2     & 1     & 3     & 2     & 2     & 1     & 3     & 2  \\ \hline
				bitops-bits-in-byte 			& 2     & 1     & 2     & 1     & 3     & 2     & 2     & 1     & 3     & 2  \\ \hline
				bitops-bitwise-and 				& 0     & 0     & 0     & 0     & 0     & 0     & 0     & 0     & 0     & 0  \\ \hline
				bitops-nsieve-bits 				& 3     & 2     & 3     & 2     & 3     & 2     & 3     & 2     & 3     & 2  \\ \hline
				controlflow-recursive 		& 4     & 6     & 4     & 3     & 4     & 6     & 4     & 6     & 4     & 6  \\ \hline
				crypto-aes 								& 17    & 16    & 17    & 16    & 13    & 16    & 17    & 16    & 13    & 14 \\ \hline
				crypto-md5 								& 21    & 30    & 21    & 30    & 3     & 2     & 21    & 30    & 12    & 15 \\ \hline
				crypto-sha1 							& 18    & 23    & 18    & 23    & 3     & 2     & 18    & 23    & 9     & 8  \\ \hline
				date-format-tofte 				& 18    & 18    & 19    & 20    & 2     & 1     & 3     & 2     & 3     & 2  \\ \hline
				date-format-xparb 				& 0     & 0     & 14    & 14    & 13    & 17    & 14    & 14    & 5     & 5  \\ \hline
				math-cordic 							& 5     & 5     & 5     & 5     & 5     & 5     & 5     & 5     & 5     & 5  \\ \hline
				math-partial-sums 				& 2     & 1     & 2     & 1     & 2     & 1     & 2     & 1     & 2     & 1  \\ \hline
				math-spectral-norm 				& 6     & 6     & 6     & 6     & 6     & 6     & 6     & 6     & 6     & 6  \\ \hline
				regexp-dna 								& 0     & 0     & 0     & 0     & 0     & 0     & 0     & 0     & 0     & 0  \\ \hline
				string-base64 						& 3     & 2     & 3     & 2     & 3     & 2     & 3     & 2     & 3     & 2  \\ \hline
				string-fasta 							& 5     & 4     & 5     & 4     & 5     & 4     & 5     & 4     & 5     & 4  \\ \hline
				string-tagcloud 					& 4     & 4     & 12    & 15    & 2     & 1     & 11    & 17    & 3     & 2  \\ \hline
				string-unpack-code 				& 0     & 0     & 5     & 4     & 5     & 8     & 12    & 64    & 5     & 20 \\ \hline
				string-validate-input 		& 4     & 3     & 5     & 4     & 5     & 4     & 5     & 4     & 5     & 4  \\ \hline
        \textbf{$\sum$}           & \textbf{169} 	   & \textbf{192}    & \textbf{209} & \textbf{235} & \textbf{135} & \textbf{146} & \textbf{197} & \textbf{284} & \textbf{155} & \textbf{186} \\ \hline
    \end{tabular}
    \label{table:stats}
		\vspace{-15pt}
\end{table*}

\vspace{-2pt}
\subsection{Manual evaluation}
\label{sec:evaluation}
\vspace{-2pt}
As part of the qualitative analysis of the results, we evaluated all the 348 call edges found by the five tools on the 26 SunSpider benchmark programs.
The manual evaluation was performed by two of the authors by going through all the edges in the merged JSON files and looking at the JavaScript sources to identify the validity of those edges.
As an output, each edge of the JSON has been extended with a ``valid'' flag that is either \emph{true} or \emph{false}.
After both authors evaluated the edges, they compared their validation results and resolved those two cases where they disagreed initially.
Upon consensus, the final validated JSON has been created.
% that is made public in the online appendix package.

As for the Node.js modules, the large number of edges made it impossible to validate all of them.
In this case we selected a statistically significant representative random sample of edges (see Section \ref{sec:qualitative} for the numbers) to achieve a 95\% confidence level with a 5\% margin of error.
One of the authors of the paper manually checked all these selected edges in the Node.js sources.

\vspace{-5pt}
\subsection{Performance measurement}
\vspace{-2pt}

%All the tools were really easy to install and use.
%If the basic environment (Java for the Java-based programs; Node.js and npm for the Node.js-based programs) is installed on the computer, these tools can be used almost instantly.
We ran the tools on an average personal computer with Windows 7.
The main hardware characteristics were Intel Core i7-3770 processor (at 3.90 GHz), 16 Gb RAM, and 1 Tb HDD (7200 rpm).
We note that besides TAJS (which can measure the time of its analysis phases), neither of the tools can measure their own running time and/or memory usage.

To measure the memory usage of the tools uniformly, we implemented a small tool which queries the operating system's memory usage at regular intervals and stores the acquired data for each process.
%Therefore we can measure the memory usage of all the tools in the same way.
In order to acquire running time data, we modified each tool's source code.
For the two Node.js tools (ACG and npm callgraph), we used the \texttt{process.hrtime()} method to calculate running time.
We also had to set the maximum heap size to 6 Gb.

For the three Java-based tools (WALA, Closure Compiler, and TAJS), we set the maximum heap size to 11 Gb.
%However, it is not necessarily precise as there can be garbage collecting while executing the call graph building algorithm and the Java runtime also needs some resources, so we used the data only for getting an estimation of memory usage.
%We set the maximum heap size to 11 Gigabytes which was enough for all cases.
For running time measurement, we used timestamps from the \texttt{System.nanoTime()} method.
\vspace{-5pt}
\section{Results}
\label{sec:results}
\vspace{-3pt}

\subsection{Quantitative analysis}
\textbf{SunSpider benchmark results.}
To evaluate the basic capabilities of the selected tools, we used the SunSpider benchmark for the WebKit browser engine (i.e. the first test program group).
This package contains 26 files which we analyzed one at a time with each tool.
After the analysis, we collected the different outputs and we converted them to our previously defined JSON format (see Section~\ref{sec:methodology-overview}).
We calculated some basic statistics from the gathered data that can be seen in Table~\ref{table:stats}.
The table shows the number of nodes (functions) and edges (possible calls between two functions) found by each tool for every benchmark program.
As can be seen, there are programs for which the number of nodes and edges are the same for all tools (e.g. bitops-bitwise-and.js, math-partial-sums.js).
There are also programs for which the results are very close, but not exactly the same (e.g. bitops-3bit-bits-in-byte.js, string-validate-input.js) and consensus could be made easily.
We should note, however, that tools produce similar call graphs typically for small programs with only a few functions, where there is only a small room for disagreement.
Finally, there are programs where the numbers show a relatively large variance across the call graph tools (e.g. 3d-raytrace.js, date-format-tofte.js).

\textbf{Node.js module results.}
To evaluate the practical capabilities of the selected tools, we chose six real-world, popular open-source Node.js modules.
Details about the subject programs can be found in Section~\ref{sec:reallife_node}.

Unfortunately, npm callgraph and WALA were unable to analyze whole, multi-file projects because they cannot resolve calls among different files (e.g., requiring a module).
TAJS supports the require command, nonetheless it was still unable to detect call edges in multi-file Node.js projects.
On the other hand, Closure Compiler and ACG were able to recognize these kind of calls.
Thus, we used only these two tools to perform the analysis on the selected Node.js modules.

%We selected 12 popular Node.js modules based on our criteria described in Section~\ref{sec:reallife_node}.
%However, ACG uses an older version of the \emph{esprima} parser, which failed to parse some of the projects.
%The parsing typically halted with the following two errors, ``Unexpected reserved word'' and ``Unexpected token''.
%These exceptions are both come from the parser itself.
%We tried to fix these problems with a parser update, but it was not enough to change \emph{esprima}, a deeper upgrade would have been required that we could not perform.
%Although Closure Compiler could analyze all the projects, we kept only those 6 that were successfully parsed by both tools.

\begin{table}[ht!]
\centering
\scriptsize
\vspace{-9pt}
\caption{Node.js results}
\vspace{-7pt}
\begin{tabular}{|l|l|l|l|l|}
\hline
            & \multicolumn{2}{c|}{ACG}  & \multicolumn{2}{c|}{Closure Compiler} \\ \hline
file        & nodes        & edges         & nodes             & edges             \\ \hline
debug       & 19           & 15            & 12                & 8                 \\ \hline
doctrine    & 85           & 175           & 53                & 174               \\ \hline
express     & 82           & 186           & 118               & 239               \\ \hline
jshint      & 349          & 1001          & 320               & 1236              \\ \hline
passport    & 41           & 40            & 39                & 49                \\ \hline
request     & 122          & 223           & 123               & 239               \\ \hline
\textbf{$\sum$}   & \textbf{698} & \textbf{1640} & \textbf{665}      & \textbf{1945}     \\ \hline
\end{tabular}
\vspace{-9pt}
\label{table:node_stats}
\end{table}

We calculated some basic statistics from the gathered data that is shown in Table~\ref{table:node_stats}.
The table displays the number of nodes (functions) and edges (possible calls between two functions) found by the tools.
As can be seen, the results show resemblance, the correlation between nodes and edges found by the two tools is high.
%there are not a single program for which the number of nodes and/or edges are the same for both tools.
However, not surprisingly, there are no exact matches in the number of nodes and edges for such complex input programs.
%programs for which the results are close, but not the same (e.g. request, jshint).
%And there are program for which one part of the results are really close, but the other part is not even close. 
It is interesting though, that for doctrine the number of edges found by ACG and Closure Compiler is very close (175 and 174, respectively), but there is a large difference in the number of nodes found by the tools.
%Nonetheless, there were no completely contradictory results reported by the tools.

%%%%%%%%%%%%%%%%%%%%%%%%%%%%%%%%%%%%%%%%%%%%%%%%%%%%%%%%%%%%%%%%%%%%%
\vspace{-5pt}
\subsection{Qualitative analysis}
\label{sec:qualitative}
For qualitatively comparing the results, we applied our exact line information based call graph comparison tool described in Section~\ref{sec:graph-compare}.
With this, we could identify which call edges were found by the various tools and compare the amount of common edges by all tools, or the edges found by only a sub-set of the tools.

\smallskip
\textbf{SunSpider benchmark results.}
First, we present the qualitative analysis on the single file SunSpider JavaScript benchmark programs.
%The diagrams below are created by the Venny~\cite{citeulike:6994833} diagram creation tool.
All the Venn diagrams are available in an interactive version as well in the online appendix package, where one can query the list of edges belonging to each area.

\begin{figure}
\centering
\includegraphics[width=\columnwidth-10mm]{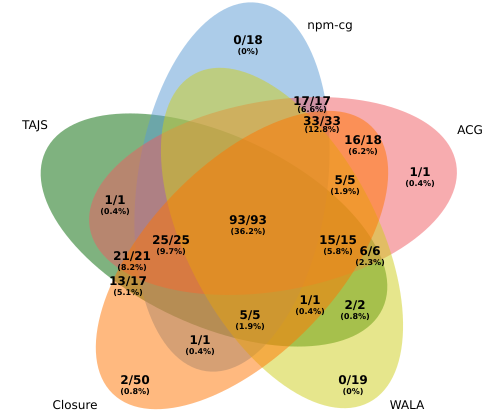}
\caption{Venn diagram of the true/total number of edges found by the tools}
\label{fig:venn_total}
\vspace{-17pt}
\end{figure}

%\subsubsection{Raw call graph extraction results}
Figure~\ref{fig:venn_total} presents the Venn diagram of the found call edges in the total of 26 benchmark programs by the five tools.
The first numbers show the true edges according to our manual evaluation (see Section~\ref{sec:evaluation}), while the second numbers are the amount of total edges.
The percentages below the two numbers display the ratio of true edges in that area compared to the total number of true edges found by the tools (which is 257 out of 348).
This representation highlights the number of edges found by all possible sub-sets of the five tools.
%The darker background color of an area, the larger amount of edges fall into that category.

In total, 93 edges were found by all the five subject tools, all of them being true positive calls.
However, four of the tools found edges that the others missed.
Although WALA, Closure Compiler and npm callgraph (npm-cg) reported a significant amount of edges that no other tools recognized, most of them turned out to be false positives during their manual evaluation.
%However, these cases are very interesting, thus we examined all the edges found only by a single tool manually.

\paragraph{Edges found by npm-cg only}
The manual analysis of the 18 unique edges found only by the npm-cg tool showed that all of these are false positive edges.
Every edge represents a call from the global scope of the program to a function.
Even though the reported calls exist, all of the call sites are within another function and not in the global scope.
Listing~\ref{lst:js1} shows a concrete example\footnote{toplevel:1:1->access-nbody.js:74:1} from the \emph{access-nbody.js} benchmark program.
The tool reports a call of Sun() (line 1) from the global scope, but it is called within an anonymous function (line 8) from line 10.
This call is properly recognized by all the other tools, however.

\begin{lstlisting}[caption={A false call edge found by npm-cg},language=json,firstnumber=1,label=lst:js1,captionpos=b,basicstyle=\scriptsize,xleftmargin=3.5ex]
function Sun(){
  return new Body(0.0, 0.0, 0.0, 0.0, 0.0, 0.0, SOLAR_MASS);
}
...
var ret = 0;

for ( var n = 3; n <= 24; n *= 2 ) {
  (function(){
    var bodies = new NBodySystem( Array(
      Sun(),Jupiter(),Saturn(),Uranus(),Neptune()
    ));
   ...
  }
}
\end{lstlisting}
\vspace{-5pt}

\paragraph{Edge found by ACG only}
There is only one edge found by ACG and no one else, which is true positive.
It is a call\footnote{date-format-tofte.js:186:5->date-format-tofte.js:8:29} to a function added to the built-in \emph{Date} object via its prototype property in \emph{date-format-tofte.js}.
Listing~\ref{lst:js2} shows the excerpt of this call.
%Both of them are valid, true positive edges.
%Moreover, both are realized through quite complex control flow that none of the other tools were able to recognize.
%Listing~\ref{lst:js2} shows the excerpt of such a possible call\footnote{3d-raytrace.js:189:29->3d-raytrace.js:357:23} in the 3d-raytrace.js program.
%Here, a function defined in another function stored in \emph{floorShader} variable gets into a \emph{shader} property of an element in the global \emph{triangles} array (line 20), which is then read and called by the \emph{intersect} function (line 10).

\paragraph{Edges found by WALA only}
In case of WALA, all the 19 unique edges are false, but for different reasons.
%There are three main categories of edges, two of them contains false, while one contains true edges.

5 of the 19 edges have a target function of ``unknown'', thus WALA was not able to retrieve the target node of the call edge.
We checked these instances manually and found that all these unknown nodes are implied by \emph{Array()} calls.
As all the built-ins and external calls are omitted from the analysis, these edges are clearly false ones.

\begin{lstlisting}[caption={A true call edge found by ACG},language=json,firstnumber=1,label=lst:js2,captionpos=b,basicstyle=\scriptsize,xleftmargin=3.5ex]
Date.prototype.formatDate = function (input,time) {
  ...
  function W() {
    ...
    var prevNY = new Date("December 31 " + (Y()-1) + " 00:00:00");
    return prevNY.formatDate("W");
  }
  ...
}
\end{lstlisting}
\vspace{-5pt}

Another group of ten false edges come from the \emph{date-format-xparb.js} program.
This program contains a large switch-case statement that builds up calls to various functions as strings.
These dynamically created strings are then executed using the \emph{eval()} command to extend the prototype of \emph{Date} object with generated formatting functions.
These formatting functions are then called from a function named \emph{dateFormat}.
WALA recognizes direct edges from \emph{dateFormat} to the functions generated into the body of the formatting functions, which is invalid, as the functions are called from the dynamically created formatting functions that are called by \emph{dateFormat}.

The last four false edges are due to invalid recursive call edges reported in the \emph{string-unpack-code.js} program.
There are several functions identified by the same name in different scopes, but WALA was unable to differentiate them.
%There are two more edges found by WALA, which are true positives contrary to the above two groups.
%One is located in the bitops-3bit-bits-in-byte.js and the other one is in the bitops-bits-in-byte.js.
%Both represent call edges where the target function is passed via a parameter to the caller function.
%None of the other tools recognized these type of call edges.

\paragraph{Edges found by Closure only}
Closure found a couple of recursive edges that no other tool did.
For example, Listing~\ref{lst:js4} shows an edge\footnote{string-tagcloud.js:99:37->string-tagcloud.js:99:37
} in the \emph{string-tagcloud.js} program, where the \emph{toJSONString} function is called from its body (line 8).

\begin{lstlisting}[caption={A true recursive call edge found by Closure},language=json,firstnumber=1,label=lst:js4,captionpos=b,basicstyle=\scriptsize,xleftmargin=3.5ex]
Object.prototype.toJSONString = function (w) {
...
    switch (typeof v) {
      case 'object':
        if (v) {
          if (typeof v.toJSONString === 'function') {
            a.push(k.toJSONString() + ':' +
               v.toJSONString(w));
        }
\end{lstlisting}
\vspace{-2pt}

48 out of the 50 unique edges by Closure is in the \emph{string-unpack-code.js} program.
All of them are false positive edges.
The reason is that Closure seems to ignore the visibility of identifiers within scopes (similarly to that observed in case of WALA).
Listing~\ref{lst:js5} shows a sketch of the problematic calls.

\begin{lstlisting}[caption={A confusing code part from string-unpack-code.js},language=json,firstnumber=1,label=lst:js5,captionpos=b,basicstyle=\scriptsize,xleftmargin=3.5ex]
var decompressedMochiKit = function(p,a,c
    ,k,e,d){e=function(c){return(c<a?"":
    e(parseInt(c/a)))+((c=c%a)>35?String.
    fromCharCode(c+29):c.toString(36))}
    ...
}(...);
var decompressedDojo = function(p,a,c
    ,k,e,d){e=function(c){return(c<a?"":
    e(parseInt(c/a)))+((c=c%a)>35?String.
    fromCharCode(c+29):c.toString(36))}
    ...
}(...);
\end{lstlisting}
\vspace{-2pt}

The inner function redefining parameter \emph{e} of the outer function (line 2) is called within itself (line 3), which is correctly identified by Closure and TAJS, but no other tools.
However, Closure reports edges from the same location to all the other places where a function \emph{e} is called (e.g. line 9), which is false, because that \emph{e} is not the same \emph{e} as it is already in another body block referring to another locally created function denoted by \emph{e}.
The \emph{string-unpack-code.js} defines four deeply embedded functions with the same parameter names, hence most of the found edges are false.

\paragraph{Interesting edges found by TAJS}
TAJS did not find any edges that were missed by all the other tools.
However, it did find some interesting edges detected only by one other tool.
One such call edge is through a complex control flow that was missed by the tools except for TAJS and ACG.

Moreover, TAJS was the only tool besides WALA that detected edges coming from higher-order function calls. %, the type of edges found only by WALA besides TAJS.
Listing ~\ref{lst:js3} shows such a call\footnote{\vspace{-10pt}bitops-3bit-bits-in-byte.js:28:1->bitops-3bit-bits-in-byte.js:7:1} in \emph{bitops-3bit-bits-in-byte.js}.

%Listing~\ref{lst:js3} displays an example\footnote{bitops-3bit-bits-in-byte.js:28:1->bitops-3bit-bits-in-byte.js:7:1} from bitops-3bit-bits-in-byte.js.

\begin{lstlisting}[caption={A true call edge found by WALA and TAJS},language=json,firstnumber=1,label=lst:js3,captionpos=b,basicstyle=\scriptsize,xleftmargin=3.5ex]
function fast3bitlookup(b) {
  ...
}
...
function TimeFunc(func) {
  ...
  for(var y=0; y<256; y++) sum += func(y);
  ...
}
sum = TimeFunc(fast3bitlookup);
\end{lstlisting}
\vspace{-4pt}

%\subsubsection{Validated and combined tool results}

%\begin{figure}
%\centering
%\includegraphics[width=\columnwidth-10mm]{fig/jVenn_chart_combined-no0.png}
%\caption{Venn diagram of the true/total number of edges found by the tools}
%\label{fig:venn_true}
%\vspace{-17pt}
%\end{figure}

As we systematically evaluated all the 348 found call edges,
%We were also interested in what are the true positive edge ratios in the various intersections of the Venn's diagram presented in Figure~\ref{fig:venn_total}.
%Moreover, knowing precisely which edges are valid, 
we could also calculate the well-known information retrieval metrics (precision and recall) for each tool and their arbitrary combinations.
We would like to note, however, that evaluation and comparison was done for simple call edges; paths along these edges were not taken into consideration.
Missing or extra edges might have different impact depending on the number of paths that go through them, thus precision and recall values might be different for the found call chain paths.

%Figure~\ref{fig:venn_total} contains the modified version of the original Venn's diagram, where the first number in each area represents the number of true positive edges, while the second number is the amount of total edges (i.e. the same numbers as shown in Figure~\ref{fig:venn_total}).
%The percentages below the two numbers display the ratio of true edges in that area compared to the total number of true edges found by the tools (which is 257).

% Table generated by Excel2LaTeX from sheet 'Munka1'
\begin{table}[htbp]
  \centering
	\scriptsize
  \caption{precision and recall measures for tools and combinations}
	\vspace{-7pt}
    \begin{tabular}{|l|r|r|r|r|r|r|}
    \hline
    Tool & \multicolumn{1}{l|}{TP} & \multicolumn{1}{l|}{All} & \multicolumn{1}{l|}{TP$^{*}$} & \multicolumn{1}{l|}{Prec.} & \multicolumn{1}{l|}{Rec.$^*$} & \multicolumn{1}{l|}{F} \\
    \hline
    npm-cg & 174   & 192   & 257   & 91\% & 68\% & 77\% \\
    \hline
    ACG   & 233   & 235   & 257   & 99\% & 91\% & 95\% \\
    \hline
    WALA  & 127   & 146   & 257   & 87\% & 49\% & 63\% \\
    \hline
    Closure & 230   & 284   & 257   & 81\% & 89\% & 85\% \\
    \hline
		TAJS & 182   & 186   & 257   & 98\% & 71\% & 82\% \\
    \hline
    npm-cg+ACG & 239   & 259   & 257   & 92\% & 93\% & 93\% \\
    \hline
    npm-cg+WALA & 203   & 219   & 257   & 93\% & 79\% & 85\% \\
    \hline
    npm-cg+Closure & 247   & 319   & 257   & 77\% & 96\% & 86\% \\
    \hline
		npm-cg+TAJS &  233  &  255  & 257   & 91\% & 91\% & 91\% \\
    \hline
    ACG+WALA & 241   & 262   & 257   & 92\% & 94\% & 93\% \\
    \hline
    ACG+Closure & 255   & 309   & 257   & 83\% & 99\% & 90\% \\
    \hline
		ACG+TAJS &  254  &  260  & 257   & 98\% & 99\% & 98\% \\
    \hline
    WALA+Closure & 238   & 311   & 257   & 77\% & 93\% & 84\% \\
    \hline
		WALA+TAJS &  187  &  210  & 257   & 89\% & 73\% & 80\% \\
    \hline
		Closure+TAJS &  239  &  293  & 257   & 82\% & 93\% & 87\% \\
    \hline
    npm-cg+ACG+WALA & 242   & 281   & 257   & 86\% & 94\% & 90\% \\
    \hline
    npm-cg+ACG+Closure & 255   & 327   & 257   & 78\% & 99\% & 87\% \\
    \hline
		npm-cg+ACG+TAJS &  255  &  279  &  257  & 91\% & 99\% & 95\% \\
    \hline
    npm-cg+WALA+Closure & 255   & 346   & 257   & 74\% & 99\% & 85\% \\
    \hline
		npm-cg+WALA+TAJS &  238  &  258  &  257  & 92\% & 93\% & 92\% \\
    \hline
		npm-cg+Closure+TAJS &  256  &  328  &  257  & 78\% & 99\% & 88\% \\
    \hline
    ACG+WALA+Closure & 257   & 330   & 257   & 78\% & 100\% & 88\% \\
    \hline
		ACG+WALA+TAJS &  254  &  279  &  257  & 91\% & 99\% & 95\% \\
    \hline
		ACG+Closure+TAJS &  257  &  311  &  257  & 83\% & 100\% & 90\% \\
    \hline
		WALA+Closure+TAJS &  239  &  312  &  257  & 77\% & 93\% & 84\% \\
    \hline
    npm-cg+ACG+WALA+Closure & 257   & 348   & 257   & 74\% & 100\% & 85\% \\
    \hline
		npm-cg+ACG+WALA+TAJS &  255  &  298  &  257  & 86\% & 99\% & 92\% \\
    \hline
		npm-cg+ACG+TAJS+Closure &  257  &  329  &  257  & 78\% & 100\% & 88\% \\
    \hline
		npm-cg+TAJS+WALA+Closure &  256  &  347  &  257  & 74\% & 99\% & 85\% \\
    \hline
		TAJS+ACG+WALA+Closure &  257  &  330  &  257  & 78\% & 100\% & 88\% \\
    \hline
		ALL &  257  &  348  &  257  & 74\% & 100\% & 85\% \\
    \hline
    \end{tabular}%
  \label{tab:tool-prec}%
	\vspace{-18pt}
\end{table}%

Table~\ref{tab:tool-prec} contains the detailed statistics of the tools.
The first column (Tool) is the name of the tool or combination of tools.
The second column (TP) shows the total number of true positive instances found by the appropriate tool or tool combination.
In the third column (All), we display the total number of edges found by the appropriate tool or tool combination.
Fourth column (TP$^{*}$) shows the total number of true edges as per our manual evaluation (i.e. it is 257 in each row).
The fifth (Prec.), sixth (Rec.$^*$), and seventh (F) columns contain the precision (TP / All), recall (TP$^{*}$ / TP) and F-measure values, respectively.

We must note that Rec.$^*$ is not the classical recall measure.
We did not strive to discover all possible call edges during manual validation, rather simply checked whether an edge reported by a tool is true or not.
Thus we used the union of all true edges found by the five tools as our golden standard.
This is just a heuristic but it provides a good insight into the actual performances of the tools compared to each other.

From the individual tools, ACG stands out with its almost perfect (99\%) precision and quite high recall (91\%) values.
While TAJS and npm-cg maintain similarly high precision (98\% and 91\%, respectively), their recall (71\% and 68\%) are far below ACG's.
Closure's recall (89\%) is very close to that of ACG, but it has significantly lower precision (81\%). 
WALA has a moderate precision (87\%), but the worst recall (49\%) in our benchmark test.

Looking at the two tool combinations, ACG+TAJS stand out based on F-measure, together they perform almost perfectly (98\% precision and 99\% recall).
It looks like they complement each other quite well.
In fact, they seem to be a perfect combination as there are no other three, four or five tool combinations that would even come close to this F-measure score.
ACG, TAJS, and Closure reach the maximum recall together, while maintaining a precision of 83\%.
Taking all the tools into consideration, the combined precision decreases to 74\% with a perfect recall.

\smallskip

\textbf{Node.js module results.}
As we already described, only ACG and Closure were able to analyze the state-of-the-art Node.js modules listed in Table~\ref{tab:node-examples}.
%Thus, Figure~\ref{fig:venn_node} shows a Venn's diagram with only two sets of edges.
From the 2281 edges found together by the two tools in the six modules, 1304 are common, which is almost 60\%.
It is quite a high number considering the complexity of Node modules coming from structures, like event callbacks, module exports, requires, etc.
There were 336 edges (14.7\%) found only by ACG and 641 (28.1\%) found only by Closure.

As the amount of edges here is an order of magnitude larger than in the case of SunSpider benchmarks, we were not able to entirely validate the found calls manually.
However, we evaluated a statistically significant amount of random samples.
To achieve a 95\% confidence level with a 5\% margin of error, we evaluated 179 edges that were uniquely found by ACG, 240 edges from those found only by Closure, and 297 from the common edges.
We found that 149 out of 179 (83.24\%) edges were true for ACG, 40 out of 240 (16.66\%) edges were true for Closure, and 248 out of 297 (83.5\%) were true for the common edges.

%\begin{figure}
%\centering
%\includegraphics[width=\columnwidth-15mm]{fig/node_venn.png}
%\caption{Venn's diagram of the edges found by acg and closure on node modules}
%\label{fig:venn_node}
%\end{figure}

%%%%%%%%%%%%%%%%%%%%%%%%%%%%%%%%%%%%%%%%%%%%%%%%%%%%%%%%%%%%%%%%%%%%%
\vspace{-5pt}
\subsection{Performance analysis}
\vspace{-3pt}
%A nagy grafok epitesenek ideje, memorialimit. Javas cuccok eseteben es csak egy valamilyen kozelites, JVM 6 G volt aladva, kiveve az utolsot, mert ott 12 G, 9G-nel meg lehalt.

In this section, we present the results of the performance testing. 
We would like to note that the measurement results contain every step of call graph building, including reading the input files and writing the output.
That was necessary because different tools implement call graph building in different ways, but reading input and writing output is a common point.
%And even these common points can be the bottleneck if they are implemented poorly (e.g. using simple read file method instead of using read stream from file).
We ran the tools ten times on each of the generated inputs and used the averages as a result (see Table~\ref{table:stats_perf}).
We highlighted the best runtime and memory consumption in each line.

\begin{table*}[ht!]
    \centering
		\vspace{-5pt}
		\caption{Performance measurements (memory in megabytes, runtime in seconds)}
		\vspace{-5pt}
		\scriptsize
    \begin{tabular}{|l|l|l|l|l|l|l|l|l|l|l|l|}
        \hline
        & & \multicolumn{2}{c|}{npm callgraph} & \multicolumn{2}{c|}{ACG} & \multicolumn{2}{c|}{WALA}      & \multicolumn{2}{c|}{Closure Compiler} & \multicolumn{2}{c|}{TAJS} \\ \hline
        Category & File      & Memory         & Runtime          & Memory       & Runtime      & Memory           & Runtime     & Memory                & Runtime    & Memory  & Runtime    \\ \hline% \hline
        \multirow{3}{*}{Simple}
            & s\_small.js  & 404  & 13.33   & \textbf{237}   & \textbf{3.11} & 1151  & 16.55    & 519   & 6.41  & 718   & 5.18    \\ \cline{2-12}
            & s\_medium.js & 2234 & 175.76  & \textbf{1168}  & 49.35  & 2537  & 181.62   & 1338  & \textbf{17.28} & 1671  & 23.83    \\ \cline{2-12}
            & s\_large.js  & 5702 & 1401.88 & 3338  & 636.49 & 8784.22  & 1085  & 3277  & \textbf{50.16} & \textbf{3132}  & 102.91    \\ \hline %\hline
        \multirow{2}{*}{Complex} 
            & c\_medium.js  & 281  & 4.76  & \textbf{239} & \textbf{2.56} & 826  & 8.27   &         411   & 4.92  & 370  & 2.74    \\ \cline{2-12}
            & c\_large.js   & 3283 & 76.49 & 1452         & 39.63         & 4010 & 210.45 & \textbf{1388} & 27.29 & 2067 & \textbf{23.79}  \\ \hline

        %\multirow{2}{*}{Complex} & small.js  &mini.js   & 5702.91 Mb      & 1401.88 sec      & 3338.73 Mb   & 636.49 sec   & 8784.22 Mb & 1085.74 sec & 479.36 Mb       & 9.57 sec   & Mb  & sec   \\ \hline
        %small.js  & 5702.91 Mb      & 1401.88 sec      & 3338.73 Mb   & 636.49 sec   & 8784.22 Mb & 1085.74 sec & 479.36 Mb       & 9.57 sec   & Mb  & sec   \\ \hline
    \end{tabular}    
    \label{table:stats_perf}
		\vspace{-16pt}
\end{table*}

In general, Closure, ACG and TAJS performed best in all cases.
%It is easier to evaluate by categories as Closure Compiler is the absolute winner if we only consider the \textit{simple} category.
%Regardless of how big the input was, it produced much better results than the rivals both in terms of memory usage and running time.
%The runner-up was TAJS closely followed by ACG, but they were much slower than Closure and also used more memory in all cases.
The npm callgraph module was generally faster than WALA.
But when it comes to large inputs, WALA was 30\% faster than npm callgraph.
On the other hand, it used more than one and a half times as much memory.
The differences may vary with the sizes of the inputs, in some cases a tool was \textasciitilde28 times faster (npm callgraph vs. Closure, s\_large), and for another input only \textasciitilde3\% better than the other tool (npm callgraph vs. WALA, s\_medium).

%We cannot proclaim an absolute winner for the \textit{complex} category where more sophisticated generating methods were used.
%In case of \textit{complex} category the results show more variance.
On the medium-sized test set in the complex category, ACG performed the best closely followed by TAJS.
On the large set, Closure used the least memory, however TAJS produced the call graph in the shortest time.
It is clearly visible that the more complex problems are considered (more similar to real-world applications) the more variance is present in the runtimes and memory consumptions.
We suppose it is due to the different inner representations the tools have to build up in order to obtain a call graph.
We conjecture that Closure and ACG keep their inner representations as simple as possible, consequently call edges are easily located by them in case of simple programs.
For complex cases, this behavior could be less effective and the more complex inner representations will pay off.

We would like to stress that these results do not say anything about the correctness and accuracy of the produced output, they are simply approximate measurement data of the memory usage and running time performances.

%To sum it up ACG and Closure performed the best in the performance measurements.
%Closure and ACG keeps it simple when dealing with call graphs.
%Their inner representation is simple and only a simple intermediate language is used for call graph construction.
%Supposedily, this is due to that Closure and ACG keep the  simple programs, however in case of more complex systems on-the-fly analysis tends to be slower and demands more resources.

%There are also large variances in differences regarding memory consumption, from as low as 30\% to as high as 2100\% depending on the inputs and call graph extraction tools.

%\todo{Tool modosito patch-ek legyenek benne az online appendixben + az acg es npm output konvertalo szkripteket + kiindulasi verziok repo url + hash}

\textbf{Discussion of the results.}
Each approach and tool has its pros and cons.
%There are various types of edges that the tools are capable of extracting at varying performance costs.
%We noticed the usual trade-offs between precision and recall, performance and precision, etc.
During this comparative study, we distilled the following statements.
\begin{itemize}
	\item Recursive calls are not handled in every tool; Closure Compiler seems to be the most mature in this respect.
	\item Edges pointing to inner functions (function in a function) declarations are not handled by every tool, e.g. WALA produces a lot of false edges because of this.
	\item Only WALA and TAJS can detect calls of function arguments (i.e., higher-order functions).\footnote{We should note, however, that according to its documentation ACG might be able to identify higher-order functions in the optimistic configuration, at the cost of lower precision.}
	\item ACG and TAJS are able to track complex control flows and detect non-trivial call edges.
	\item Closure often relies only on name-matching, which can cause false or missing edges.
	\item It seems that WALA can analyze \emph{eval()} constructions and dynamically built calls from strings to some extent.
	\item The calls from anonymous functions defined in the global scope are mistreated by npm-cg, which detects a call edge directly coming from the global scope in such cases.
	\item Closure has a superior runtime performance for very large inputs with high recall at the expense of a lower precision.
	\item ACG consumes the least memory and runs the fastest among all the tools on small to medium sized inputs.
	\item WALA and npm-cg are practically unusable for analyzing code at the scale of millions of code lines.
\end{itemize}
%a) valaki nem tudja kezelni a rekurziot
%b) valaki nem talalja meg azokat a hivasokat, amik fuggvenyen beluli fuggveny deklaraciokbol/ba mennek
%c) wala Array() hivasokra ad unknown elet
%d) wala a parameterkent atadott fuggvenyek hivasat megtalalja, mas nem
%e) az acg eleg komoly control flow kovetest tud, atteteles eleket megtalal
%f) clousure sokszor buta nevegyezeseket nez csak
%g) wala mintha elemezne az eval es stringbol osszerakott dinamikus hivasokat
%h) Closure nagy recall, kisebb precision
%i) az npm-cg a global scopeba levo nevtelen fuggvenyek altali hivasokat toplevel11-nek vette.

\vspace{-5pt}
\section{Threats to Validity}\label{sec:threats}
\vspace{-2pt}
%We performed performance measurements on different configurations, thus the results were inconsistent and hard to compare.
%We used a general PC configuration for measuring performance.
%We ran each call graph builder system 10 times each for every input.
%Because only insignificant differences were experienced in the standard deviation, we used the averages as results.
A lot of factors might have affected our measurements.
%, like Java garbage collection, or the fact that we needed to modify the source code of the tools to inject measurement commands.
Some of the tools might perform additional tasks during call graph construction, which we could not omit from the measurement.
Nonetheless, we treat our performance measurement with proper care; they are only used to assess the orders of magnitudes for memory consumption and running times.

%ECMAScript 6 is not supported fully by most of the tools thus we restricted our investigations only to ECMAScript 5 compliant subject programs for single-file analysis. % which is supported by each tool.
%However, we also ran the two capable tools on state-of-the-art ECMAScript 6 compliant Node.js modules and got similar results.

%Using Java-based call graph builders does not inject any platform dependent threat in our study.
%However, there could be different builds for different platforms (caused by the nature of NPM).
%We only evaluated different tools and performed the measurements on Windows, however there should be no difficulty to perform %such analysis on Linux or on MacOS platforms.

%We had to dig into the source code of some call graph builder tools and modify it in order to obtain the call graphs themselves.
Our modifications in the tools for call graph extraction mechanism may have introduced some inconsistencies.
However, we made only slight changes and most of them affected only the reporting of edges, thus this threat has a limited effect.

We ran all the tools with default configurations.
Various parameters might have affected the performance and precision of the tools.
Nonetheless, we do not expect the main results to be much affected by these parameters.

We might have missed some good candidate tools from the comparison.
However, the presented evaluation strategy and insights are useful regardless of this.
Nevertheless, it is always possible to replicate and extend a comparative study like this.
%Facebook's Flow is a popular and efficient tool, however it does not provide a public API for call graph construction.
%Flow is written in OCaml which is a declarative programming language.
%As Flow is written in OCaml and is quite complex, unfortunately we could not manage to customize for our needs.

Regarding the manual evaluation of the call edges, the subjectivity of evaluators is also a threat.
We tried to mitigate this by having two authors validate all the edges for the 26 SunSpider benchmark test cases.
There were preliminary disagreements in only 2 out of 348 cases between the evaluators that they could resolve in the end.
Thus, we think the possible bias due to evaluation errors is negligible.

\vspace{-4pt}
\section{Conclusions}\label{sec:conclusions}

Code analysis of JavaScript programs has gained a large momentum during the past years.
Many algorithms for vulnerability analysis, coding issue detection, or type inference rely on the call graph representation of the underlying program.
%Therefore, the precise and effective creation of call graphs is vital.

In this paper, we presented a comparative study of five state-of-the-art static algorithms 
%-- implemented by the npm call graph, IBM WALA, Google Closure Compiler, Type Analyzer for JavaScript, and Approximate Call Graph tools -- 
for building JavaScript call graphs on 26 WebKit SunSpider benchmark programs and 6 real-world Node.js modules.
Our purpose was not to declare a winner, rather to get empirical insights to the capabilities and effectiveness of the state-of-the-art static call graph extractors.

Each tool had its strengths and weaknesses.
For example, Closure recognized recursive calls and had an overall good performance both in terms of running time and memory consumption, but it introduced errors due to shallow name-matching and had a relatively low recall.
ACG tracked complex control flows to find call edges and had high precision and recall at the same time with great memory consumption and runtime, but missed higher-order function calls.
WALA had the capability to detect higher-order function calls (callbacks), but produced some edges with unknown nodes and had the lowest recall and highest memory consumption of all tools.
The npm callgraph tool had very high precision, but poor performance and found no unique true call edges.
TAJS provided very conservative results, meaning that it had almost perfect precision, but very low recall, while having a very good overall performance.

%Our study can be a good starting point for practitioners to chose an appropriate tool and for call graph algorithm developers to refine their algorithms.
It is also evident from the results that the combined power of various tools is superior to those of individual call graph extractors.
Thus, we would encourage the development of algorithms that combine these state-of-the-art approaches.
%To overcome the inherent problems of dynamism in JavaScript, probably a hybrid static and dynamic approach would perform the best.

Our future plan is to extend and replicate the presented study by adding more static tools (e.g. taking commercial tools and IDEs into consideration) as well as including some dynamic call graph extraction approaches.
%Upon collecting empirical data, we plan to design a hybrid approach to build call graphs for JavaScript.

\vspace{-5pt}
\section*{Acknowledgment}
\vspace{-4pt}
%\todoi{EFOP vagy IoLT hivatkozas}
This research was supported by the EU-funded Hungarian national grant GINOP-2.3.2-15-2016-00037 titled ``Internet of Living Things'', and the UNKP-17-4 New National Excellence Program of the Ministry of Human Capacities, Hungary.
\vspace{-5pt}
\bibliographystyle{IEEEtran}
\bibliography{bibl}

\end{document}